\documentclass{llncs}

\usepackage{makeidx}
\usepackage{graphicx}
\usepackage{cite}

\begin{document}

\pagestyle{headings}

\mainmatter

\title{Belle II iTOP Optics: Design, Construction and Performance}

\titlerunning{Belle II iTOP Optics}

\author{Boqun Wang\inst{1}, Saurabh Sandilya\inst{1}, Bilas Pal\inst{1}, Alan Schwartz\inst{1}}

\authorrunning{Boqun Wang et al.}

\institute{University of Cincinnati, Cincinnati OH 45221, USA,\\
\email{boqunwg@ucmail.uc.edu}}

\maketitle

\begin{abstract}
  The imaging-Time-of-Propogation (iTOP) counter is a new type of
  ring-imaging Cherenkov counter developed for particle identification
  at the Belle II experiment. It consists of 16 modules arranged
  azimuthally around the beam line. Each module consists of one
  mirror, one prism and two quartz bar radiators. Here we describe the
  design, acceptance test, alignment, gluing and assembly of the
  optical components. All iTOP modules have been successfully
  assembled and installed in the Belle II detector by the middle of
  2016. After installation, laser and cosmic ray data have been taken
  to test the performance of the modules. First results from these
  tests are presented. 
\end{abstract}

\section{Introduction}

The Belle II~\cite{belle2}/SuperKEKB~\cite{superkekb} experiment is an
upgrade of the Belle/KEKB experiment for searching for New Physics
(NP), which is physics beyond the Standard Model (SM). The upgraded
detector is planning to take $\sim 50$ $ab^{-1}$ of $e^+e^-$ collision
data, with a design luminosity of $8 \times 10^{35} cm^{-2} s^{-1}$.
This is about 40 times larger than the KEKB collider. To achieve such
high luminosity, the so-called nano-beam technology~\cite{nano-beam}
is used to squeeze the beam bunches significantly.

Many sub-detectors of Belle will be upgraded for Belle II. This
includes the newly designed iTOP (imaging-Time-Of-Propogation)
counter~\cite{itop-oshima, itop-akatsu, itop-oshima2, itop-enari},
which is the particle identification counter in the barrel region. It
consists of a 2.7 m long quartz optics for the radiation and
propogation of the Cherenkov light, an array of micro-channel-plate
photo-multiplier tubes (MCP-PMT)~\cite{mcppmt} for photon detection,
and wave-sampling front-end readout electronics~\cite{itop-irsx,
  itop-irsx1}. This article describes the design, construction and
performance of the optics for the iTOP counter.

\section{Detector Design}

As shown in Fig.~\ref{fig:itop}, one iTOP module consists of two bars
with the dimension as 1250 $\times$ 450 $\times$ 20 mm. At one end of
the bars is the reflection mirror with spherical surface. At the other
end is the expansion block called prism. All optics components are
made of Corning 7980 synthetic fused silica, which has a high purity
and no striae inside.

When a charged track goes through the quartz radiator, it emits
Cherenkov photons. The Cherenkov angle depends on the mass of the
particle for a given momentum, the latter is measured by the central
drift chamber (CDC). The photons are reflected by the bar surfaces and
the reflection mirror, then collected by the MCP-PMTs at the prism
end. The resolution of the photon sensors and the front end
electronics are required to be better than 50 ps, which is needed to
distinguish the time of propogation difference between Cherenkov
photons from $\pi^{\pm}$ and $K^{\pm}$ tracks.

\section{Module Construction}

The construction of the real iTOP module was started in the end of
2014 and all 17 modules, including one spare, were finished by April
2016. After testing with laser and cosmic ray, these modules were
installed on Belle II detector by May 2016.

\subsection{QA of Quartz Optics}

To achieve the high K/$\pi$ separation capability of the iTOP counter,
the optics need to have very high quality. The Cherenkov photons can
reflect hundreds of times inside the quartz radiator, so the surfaces
of the quartz bars need to be highly polished. The requirement for
surface roughness is $<$ 5 \AA r.m.s., and for flatness the
requirement is $<$ 6.3 $\mu m$. For all 34 bars needed, 30 were
produced by Zygo Corporation (USA) and 4 were produced by Okamoto
Optics Works (Japan).

\begin{figure}
  \centering
  \begin{minipage}{0.45\linewidth}
    \includegraphics[width=\textwidth]{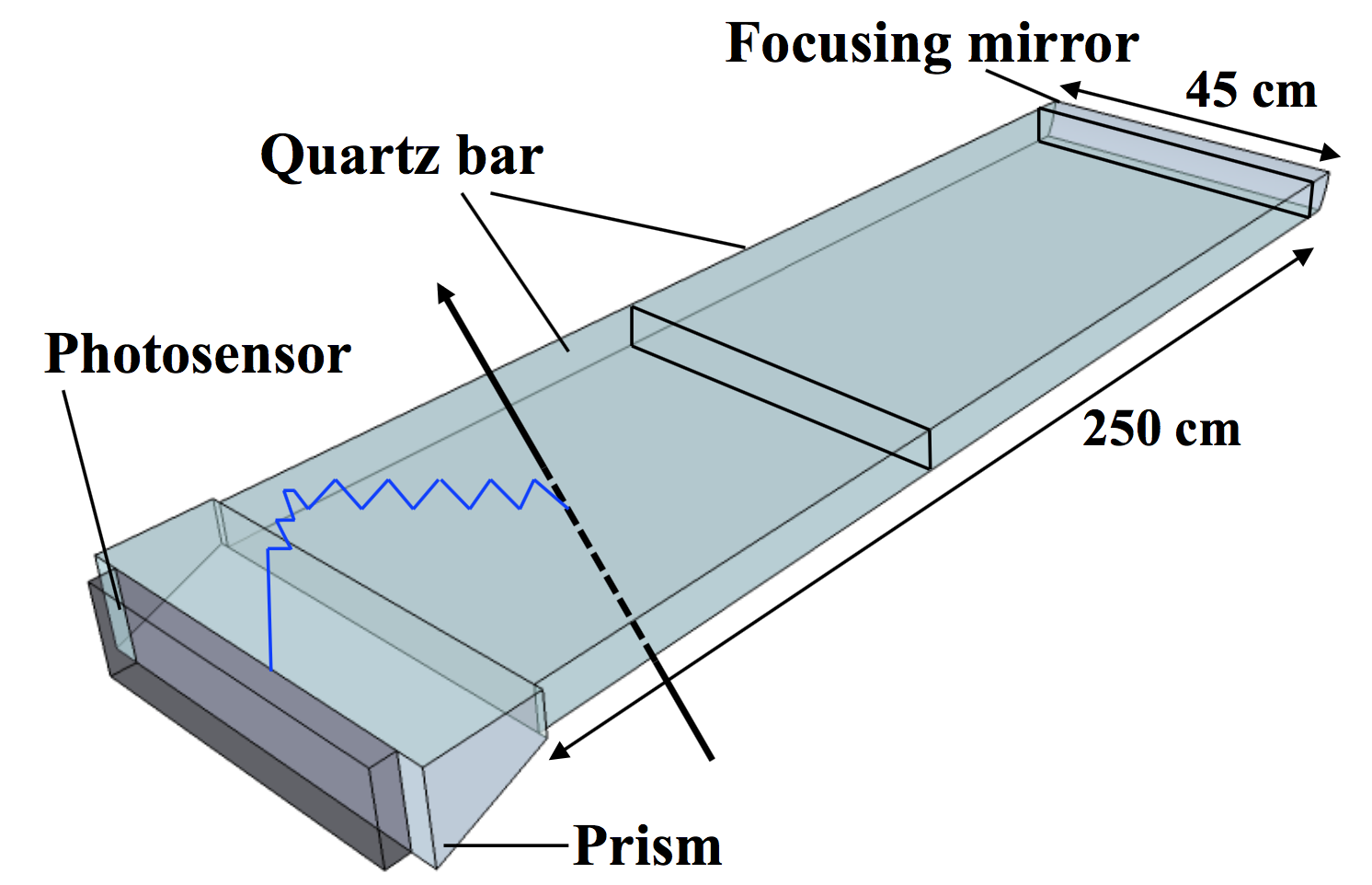}
    \caption{Optical overview of the iTOP detector.}
    \label{fig:itop}
  \end{minipage}
  \begin{minipage}{0.45\linewidth}
    \includegraphics[width=\textwidth]{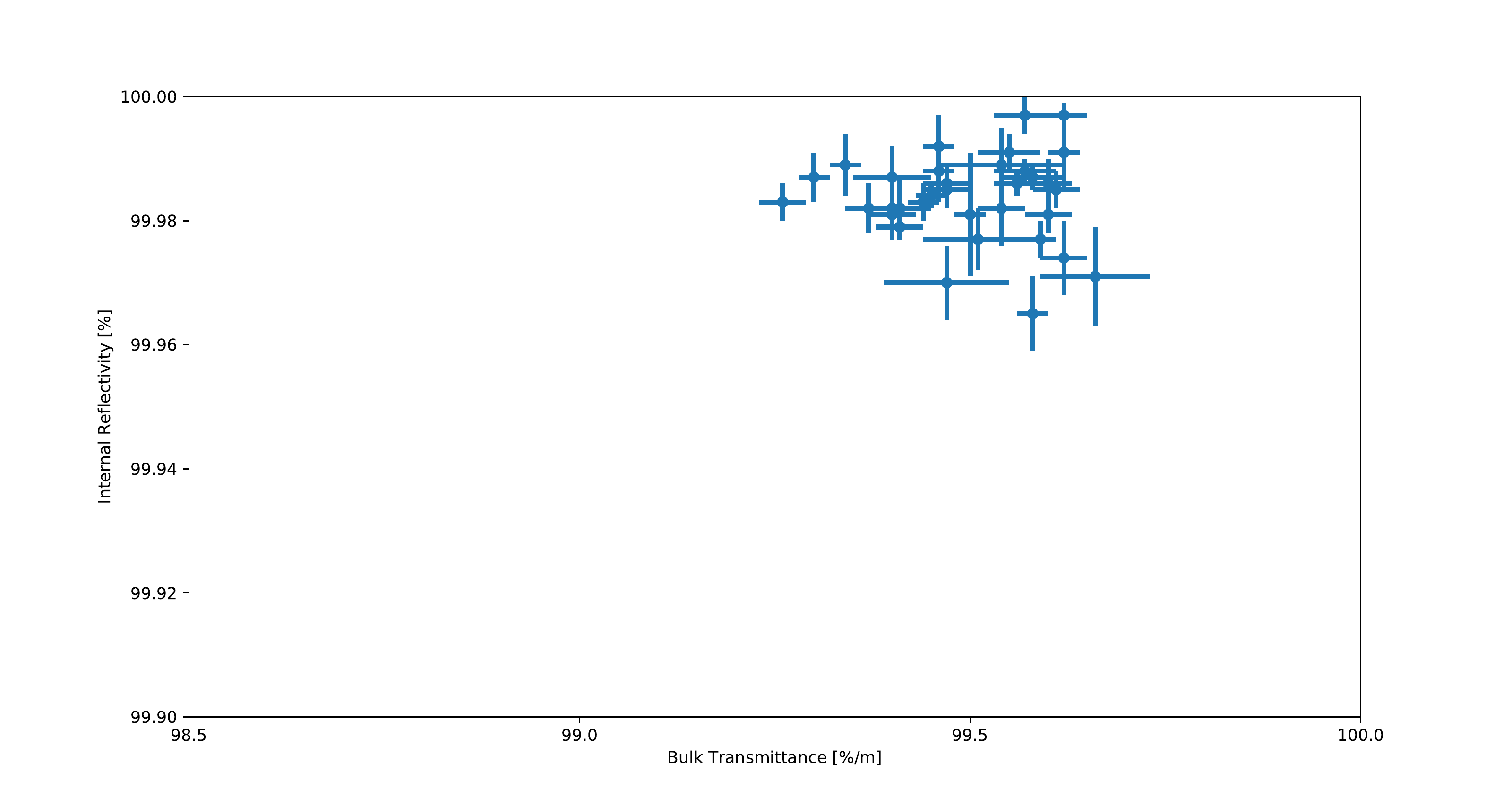}
    \caption{Summary of quartz bar QA results.}
    \label{fig:bar-qa}
  \end{minipage}
\end{figure}

After receiving the quartz bars from the vendors, they were mounted on
the measurement stage for the QA tests. By injecting laser beam
perpendicular to one surface of the bar or have a angle relative to
it, we can measure the bulk transmittance and internal reflectivity of
the quartz bar by measuring the laser intensity before and after it
went through the bar. The requirements for bulk transmittance and
internal reflectivity were $>$ 98.5 \%/m and $>$ 99.9 \%,
respectively. As shown in Fig.~\ref{fig:bar-qa}, all received quartz
bars met the requirements.

For the QA of the other optics, the most important tests were the
angle of the tilted surface of the prism and the radius of the
mirror's spherical surface. They were measured by injecting laser to
the optics and measure the laser direction after it went through or
reflected by the optics. All optics met the requirements.

\subsection{Gluing and Assembly}

After the QA process was finished and all the quartz optics passed the
requirements, two quartz bars, one prism and one mirror were mounted
on a gluing stage for precision alignment and gluing.

Two laser displacement sensors and an autocollimator were used for the
alignment. Laser displacement sensor measures the distance to the
surface. It was used to align the position, both horizontal and
vertical, of two optics' surfaces. Autocollimator injects laser to a
mirror which is mounted on the surface of the optics, and by measuring
the angle of the reflected laser relative to the original one to get
the angle information. It is used to align the angle between two
optics' surfaces.

After the alignment, the optics were moved together with a 50 $\sim$
100 $\mu m$ gap. The joints between optics were taped by using Teflon
tape to make a ``dam'' to prevent the epoxy from flowing outside. The
epoxy used for gluing was EPOTEK 301-2. It consists of two parts,
which needs to be mixed before gluing. The mixture was centrifuged to
remove the air bubbles inside. Then the adhesive was applied from a
syringe to the glue joint by using high pressure dry air. After
applying, it took 3 $\sim$ 4 days to be fully cured.

When the curing process was finished, the excessive glue was removed
using Acetone. The alignment may change after the curing, so it needed
to be measured again. The achieved horizontal and vertical angle
between the two optics near the glue joint was within $\pm$ 40 arcsec
and $\pm$ 20 arcsec, respectively.

The completed iTOP optics was moved into a Quartz Bar Box (QBB). The
QBB is a light-tight supporting structure made of aluminum honeycomb
plates. It has high rigidity with light material. The optics was
supported on by PEEK buttons that were glued to the inner surfaces of
QBB.

When the QBB assembly was completed, the MCP-PMT modules with
front-end electronics were installed at the prism end of the module.

\subsection{Installation}

After the iTOP module assembly was completed, it was transferred by
truck to the experimental hall for installation. Each module was
installed by using movable stages. A module was mounted on a guide
pipe, which was supported by the stages, so it was able to move and
rotate around the guide pipe. The module deflection during the
installation process was monitored by deflection sensors, and it was
required to be less than 0.5 mm. The installation of all the modules
was completed in May, 2016. More details can be found in
Ref.~\cite{itop-install}.

\section{Performance with Cosmic Ray}

After installation, the cosmic ray data was taken to validate detector
performance with and without the magnetic field. Six cosmic ray
triggers were prepared, with each consisted of a plastic scintilltor bar.
Their positions along the beam axis were the same with the collision
point and their positions in x-y plane can be changed as needed.

For the performance test with cosmic ray, the tracking information was
not available. The number of observed photon hits for each iTOP module
was compared between data and MC simulation. For no magnetic field
condition, the number of photon hits for data was consistent with MC
simulation within 15\%. For the 1.5T magnetic filed condition, the
discrepancy is at 20 - 30 \% level. There are many reasons for this,
including the angle and momentum distributions in the cosmic ray muon
flux, the hit identification efficiency of MCP-PMTs, etc. More
detailed studies are planned by combining tracking information from
the CDC detector.

\section{Summary}

The iTOP counter is a novel particle identification device in the
barrel region for Belle II detector. In this article we described the
design, construction and performance of the iTOP counter. The last
iTOP module has been finished and installed in May 2016, and the Belle
II detector has been moved to the beam line in April 2017. Currently
the global cosmic ray data taking is on going, for the purpose of
testing, calibration and integration of the sub-detectors including
iTOP counters.

\it{This research is supported under DOE Award DE-SC0011784. I’d like
  to thank the organizers of the TIPP 2017 conference for allowing me
  to give this talk.}

\end{document}